\documentstyle [prl,aps,epsf]{revtex}

\begin{document}

\title {``Just So" Neutrino Oscillations Are Back}

\author{Sheldon L. Glashow$^1$, Peter J. Kernan$^2$ and Lawrence M.
Krauss$^{2,3}$}
\address{
 $^1$Lyman Laboratory for Physics, Harvard University, Cambridge MA 02138\\
$^2$Department of Physics
Case Western Reserve University
Cleveland, OH~~44106-7079 \\
$^3$Also Department of Astronomy
}

\twocolumn[
\maketitle
%\tightenlines
\begin{abstract}
\widetext
\noindent
Recent evidence for oscillations of  atmospheric neutrinos at
Super-Kamiokande suggest, in the simplest see-saw interpretation, 
neutrino masses such that `just so' vacuum
oscillations can explain the solar neutrino deficit. Super-K solar
neutrino data provide preliminary support for 
this interpretation.  We describe
how the just-so signal---an energy dependent seasonal
variation of the event rate, might be detected within the coming
years and provide general arguments constraining the sign of the variation.  
The  expected variation at radiochemical detectors
may be below present sensitivity, but a significant modulation in the
$^7$Be signal could shed light on
the physics of the solar core---including a direct measure of the solar core
temperature. 
\end{abstract}

%insert suggested PACS numbers in braces on next line
\pacs{}

]

\narrowtext

A decade ago we argued \cite{GlashowKrauss} that seasonal modulation of the
solar-neutrino signal seen in various detectors could be a unique signature
of `just-so' vacuum neutrino oscillations\cite{ponte},  with the typical
solar-neutrino oscillation length comparable to the Earth-Sun distance. The
resulting oscillations could provide the large suppression needed to
explain the $^{37}$Cl data~\cite{barger,GlashowKrauss}.  Since then, much
attention has been paid to matter-enhanced (MSW) oscillations, for which 
large suppressions of the signal is obtained for a  broad range of masses,
and with which virtually any combination of suppressed signals at different
detectors is explicable. In the intervening decade, solar neutrinos have
been studied at many facilities, yet a definite resolution to the puzzle
has not been identified. We argue that the just-so scenario, although it
requires finely tuned masses, is both more easily testable and more readily
falsifiable.

The Super-Kamiokande Collaboration has announced a remarkable  result
\cite{SuperK}. Evidence for vacuum oscillations between muon neutrinos and
another  species (not $\nu_e$) was reported at a highly significant level
and with a nearly maximum mixing angle. The required squared-mass
difference is $\Delta m^2\equiv |m_1^2 -m_2^2|\simeq 0.005\;$eV$^2$.  With
the plausible hypothesis  that the oscillation involves $\nu_\mu$ and
$\nu_\tau$,  and that the mass difference is dominated by the tau neutrino
mass, this suggests $m_{\nu_{\tau}} \approx .07 $ eV.

There are several reasons to reconsider just-so solar neutrino
oscillations in light of this  result.  
While we have no
direct empirical information on the neutrino mass spectrum, an elegant yet
simple origin for the Majorana masses of
light neutrinos is the so-called
see-saw mechanism\cite{gellmramslan}, which yields:
\begin{equation}
m_{\nu_i} \approx m_{q_i,l_i}^2/M_X ,
\end{equation}
where $m_{q_i,l_i}$ are
the masses of the associated up-type quark or charged lepton in family $i$, and
$M_X$ is a heavy mass scale associated with the symmetry breaking
responsible for
neutrino masses. Taking the quark masses to be relevant,
we expect $m(\nu_\mu)\approx .07 \times (1.5/175)^2
\approx 7 \times 10^{-6}$~eV, and a much smaller $m(\nu_e)$.
These neutrino masses are just right to provide a just-so explanation
of the solar-neutrino flux.

Given our lack  of understanding the origin of  even the charged lepton
masses, one might find the above reasoning suspect.  It has been argued,
for example, that maximal vacuum oscillations with a much smaller
wavelength, resulting in a uniform halving of the observed neutrino signal, 
might resolve the solar neutrino problem, especially if there is still some
systematic error in the $^{37}$Cl  data\cite{JNB}. On the other hand  and
rather remarkably, the Super-K collaboration reports tentative evidence,
based on their measurements of the shape of the recoil electron spectrum,
for just-so oscillations with masses provisionally in the
range $ 6\le \Delta m^2 \le 60 \times 10^{-11}$ eV$^2$~\cite{SuperKK}.

If the just-so solution is correct, must we wait for confirmation from
observations of neutral-current events at SNO  or for unambiguous results
from  Borexino? We argue that this may not be necessary because of the
unique  signature of vacuum oscillations: an energy and time dependent
variation of the signal as a function of the distance to the source. 
In our
analysis, we re-examine the seasonal variation of both the overall
flux ({\it i.e.,} integrated over energy) and the recoil electron spectrum
for the mass regions now of interest.
(Note: before the most recent Super-K result several other groups 
explored various just-so signatures( i.e.
\cite{krast,lisi,smirn}) ).

Of course, all
explanations of solar neutrino deficit are subject to a small 
seasonal variation  of the overall event rate due to the eccentricity
of Earth's orbit. Additional effects characteristic of just-so solutions
are energy dependent 
 and can average out when the signal is
averaged over energies.  Thus, the {\it rate\/} measurements at Super-K and
at the radiochemical detectors, which have failed to detect seasonal
variations, cannot exclude just-so oscillations. Depending on the neutrino 
mass we shall show
that just-so oscillations can lead  to seasonal variations in
the  {\it spectrum\/} of recoil energies~\cite{GlashowKrauss}.  These
variations can be radically energy dependent and  are potentially
detectable at Super-K.  In cases where the energy
dependence of the seasonal variation is weak, the integrated event
rate can vary seasonally over 5\%\ in addition to its
expected variation due to the earth's eccentricity. This effect should soon
be detectable.   The sign of the variation may be
telling as well.  Lower masses lead to uniformly larger seasonal variations
than expected in the absence of oscillations, while larger masses can result in
variations which cancel out the variation due to the earth's eccentricity 
(although, as we derive below, this possibility is disfavored by the Super-K
spectral information).

Our neutrino mass model, motivated by Super-K data, 
involves three
left-handed mass eigenstates.  Mixing between the flavor eigenstates
is governed by three angles $\theta_1$
$\theta_2$, and $\theta_3$.  We assume the heaviest mass eigenstate,
that playing a role in atmospheric neutrino oscillations,
to have a mass of $.07$~eV. Oscillations between
a solar electron neutrino and this state, if present,
 will undergo many oscillations on their way to the detector
sun and average out to yield an essentially energy-independent
suppression. Oscillations between electron and muon
neutrinos, however, are assumed to be just-so.  Thus we take
$$m_1\sim 0,\quad m_2\sim 10^{-5}\; {\rm eV}, \quad m_3\sim 0.1\; {\rm eV}\,.$$
Adopting the conventional KM parameterization, we write the flavor
eigenstates as:
\begin{eqnarray}
&\nu_e&\equiv c_2c_3\nu_1+ c_2s_3\nu_2 + s_2e^{-i\delta}\nu_3\,,\cr
&\nu_\mu&\equiv \cr 
&&-(c_1s_3+s_1s_2c_3e^{i\delta})\nu_1
               +(c_1c_3-s_1s_2s_3e^{i\delta})\nu_2+s_1c_2\nu_3 \nonumber.
\end{eqnarray}
from which we obtain:
\begin{eqnarray}
&P(\nu_e\rightarrow\nu_e)\big\vert_{\rm solar}= &  \cr
&1-{\sin^2{2\theta_2}\over 2}-
\cos^4{\theta_2}\sin^2{2\theta_3}\sin^2{(m_2^2R/4E)}. &  \nonumber
\end{eqnarray}

For atmospheric oscillations, for which $m_2$ is too small to play
much of a role, we obtain:
\begin{eqnarray}
&P(\nu_\mu\rightarrow\nu_{\mu})\big\vert_{\rm atmospheric}= & \cr
&1-4\sin^2{\theta_1}\cos^2{\theta_2}\sin^2{(m_3^2R/4E)}
\cases{\sin^2{\theta_2} & ($\nu_e$)\cr
\cos^2{\theta_2}\cos^2{\theta_1}& ($\nu_\tau$)\cr} & \nonumber
\end{eqnarray}

Note that the CP-violating phase $\delta$ plays no role.
Super-K (and CHOOZ) suggest that $\nu_\mu\rightarrow \nu_e$ is small for
atmospheric neutrinos. In the limit $\theta_2=0$:
\begin{eqnarray}
&P\big\vert_{\rm
solar}=1-\sin^2{2\theta_3}\sin^2{(m_2^2R/4E)}, & \cr
& P\big\vert_{\rm atmospheric}=
1- \sin^2{2\theta_1}\,\sin^2{(m_3^2R/4E)}& .\nonumber
\end{eqnarray}
We use  the general form here, but 
take $\sin^2{2\theta_2} < 0.1$ to assure that atmospheric 
muon neutrinos oscillate mostly into $\nu_\tau$. 

We can readily determine the resulting energy-dependent 
seasonal suppression
of the neutrino signal  at Super-K 
as a function of neutrino mass.   
Our numerical routine uses
the B neutrino spectrum from Bahcall and Pinnsoneault \cite{bp98}\  and
standard neutrino-electron scattering cross sections.  We assume a 
sharp detector threshold of 6.5~MeV and integrate over
the Super-K electron
energy resolution function.   Predicted rates are multiplied  by a
normalization factor of
$\sim 0.9$ to account for detector efficiency,
deadtime, etc., thereby ensuring  that the observed rate of
13.5 events/day corresponds, as reported, to $\sim 47$\%\ of its 
SSM value~\cite{bp98}.

Preliminary analyses of the Super-K spectrum, and the rates at other solar
neutrino experiments, lead to a variety of estimates for parameters that
best explain the data \cite{superk,bahcall2}: Spectral measurements by
Super-K suggest  (assuming $\theta_2=0$) $\Delta m^2 \simeq 4 \times
10^{-10}$ eV$^2$ \cite{superk}. A global analysis of event rates at  all
neutrino detectors, combined with the Super-K spectrum, suggests  (again
with $\theta_2 =0$)  $\Delta m^2\simeq 7 \times
10^{-11}$~eV$^2$~\cite{bahcall2}.    Our analysis spanned both these
estimates as well as the smaller value suggested by the see-saw mechanism: 
$2\le \Delta m^2\le  60 \times 10^{-11}\; {\rm eV}^2$, with mixing angles
chosen to fit the observed rate of 13.5 events/day within uncertainties.

The predicted seasonal dependence of  the integrated Super-K signal as a
function of $\Delta m^2$ can be understood as follows:  For small mass
differences the oscillation length far exceeds the Earth-Sun distance.  As
$\Delta m^2$ increases, oscillations  kick in and the 
survival probability of an
electron neutrino  decreases monotonically, so
that as the Earth-Sun distance increases from perihelion to aphelion the
$\nu_e$ flux at the detector  decreases via the inverse-square law  {\it
and} through the effect of just-so oscillations. For  $2\le \Delta m^2 \le
6 \times 10^{-11} {\rm eV}^2$, the seasonal variation of the signal
significantly exceeds its no-oscillation value.

For $\Delta m^2\simeq 6 \times 10^{-11} {\rm eV}^2$, the mean oscillation
probability at Earth is largest and the electron neutrino flux reaches a
minimum.  For  slightly larger values, oscillations tend to {\it
increase\/} the $\nu_e$ flux as the Earth-Sun distance increases. For this
case the just-so and inverse-square effects  oppose one another and the
overall event rate can  even increase.  As $\Delta m^2$ is further increased,
the seasonal effect oscillates in sign. At the same time, the seasonal effect
becomes energy dependent so that the signals in  different energy bins at
Super-K would differ dramatically from one another. For even larger $\Delta
m^2$, the just-so signal washes out due to averaging over the different
neutrino energies: the seasonal variation becomes conventional and energy
independent.

We have explored these effects quantitatively using several
different numerical analyses. To maximize statistics and sensitivity 
to seasonal variation, the relevant
quantity to consider is not the day of the year, but rather the Earth-Sun
distance, which is probed twice
over the course of the year.   Seasonal variation should
be plotted as a function of ``days since perihelion". 
%We all know the length of the yr

Figure 1  shows the ratio $R$ of the predicted number of 
Super-K events within
90 days of perihelion to those in
the rest of the year, versus $\Delta m^2$
and with maximal oscillations. The lower part of the figure 
displays the predicted overall event rate at  Super-K versus
$\Delta m^2$.
The effects outlined earlier are evident.
For small $\Delta m^2$ the seasonal variation is enhanced
compared to the no-oscillation case. For larger values, the
seasonal variation oscillates in and out of phase with the 
$1/r^2$ variation. Moreover, the shift in the sign of the
seasonal variation (where the curve crosses the no-oscillation point)
occurs precisely at the maxima and minima of the event rate curve, 
validating the picture discussed above.   

\begin{figure}[tbp]
\epsfxsize = \hsize \epsfbox{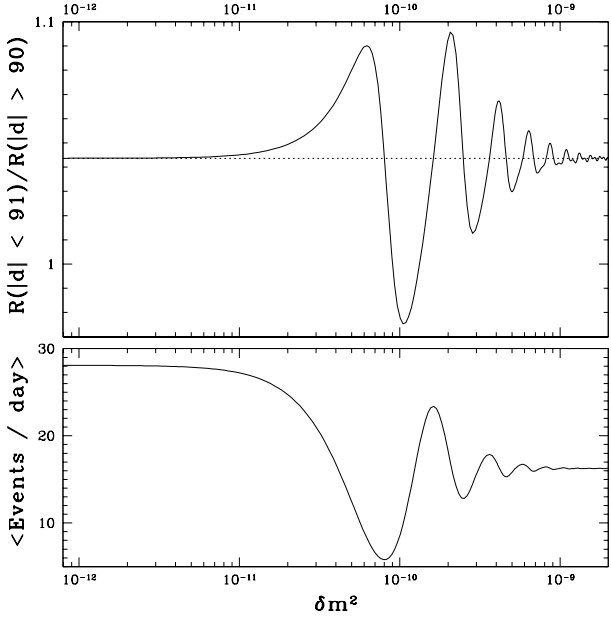}
\caption{ Upper: Seasonal variation predicted for the solar neutrino signal in
Super-K  for
maximal just-so oscillations (no oscillations---dashed curve) versus neutrino
$\Delta m^2$. The number of events within 90 days of perihelion is divided by the number of events more than 90
days removed from perihelion. Lower: Overall event rate vs. $\Delta m^2$} 
\end{figure}

Figures 2, 3, and 4 show
the time variation of  the differential and
integrated event rates at Super-K 
for three
different values of $\Delta m^2$. 
At the left, we display the temporal variation
of the predicted neutrino signal at various energies
between 6 and 14 MeV, with oscillation parameters as specified. At the
right,  for the same parameters, we display  the
temporal variation of the overall event rates.  We also indicate
the expected variation in the signal in the absence of
oscillations.

The results validate our {\it a priori\/}  reasoning and provide some new
insights.  For smaller values of $\Delta m^2$, as in fig.(2), the
integrated and differential seasonal variations are similar---as expected
because none of the detected neutrinos have completed one full 
oscillation. In this case, the overall seasonal variation can exceed its
no-oscillation value by an additional 5-6$\%$.  

For intermediate values of $\Delta m^2$,  fig.(1) shows the just-so effect 
canceling  the seasonal variation and  the event rate increasing with
$\Delta m^2$, and hence decreasing with neutrino energy. This behavior is
evident in fig.(3), where there is virtually no seasonal variation and the
event rate  drops precipitously at high energies. However, observations
at Super-K suggest an upturn in the neutrino signal at high energies. Thus,
these values of $\Delta m^2$ are disfavored. {\it The just-so
oscillations indicated by Super-K data suggest that the seasonal variation
of the overall event rate is in phase (or at least, not out of phase) 
with the inverse square
variation} (see also \cite{smirn}).   

%\vskip 3.8in
\begin{figure}[tbp]
\epsfxsize = \hsize \epsfbox{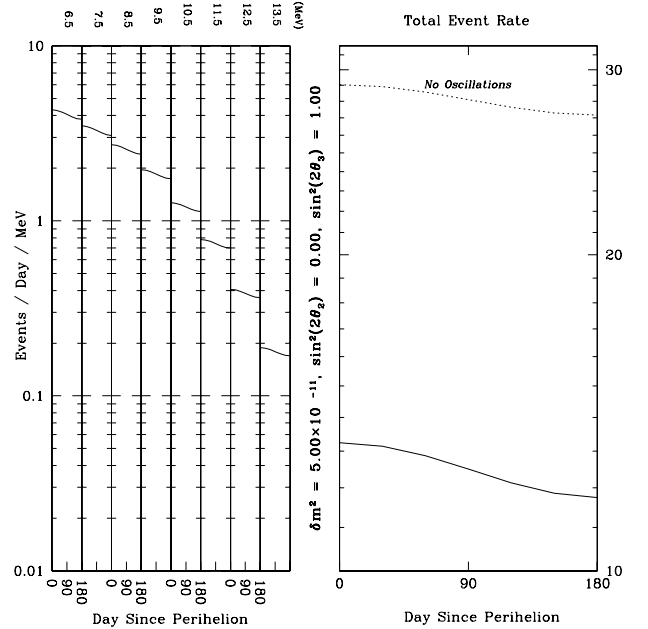}
\caption{ Predicted seasonal variation for the solar neutrino signal in
Super-K as a result of
just-so oscillations, for $\Delta m^2 = 5 \times 10^{-11}$ eV$^2$, and mixing
angle chosen to reproduce the Super-K observed event rate (within quoted
uncertainties).   On the left is shown the variation of
the number of events/day/MeV predicted at various electron energies. 
On the right is the variation  of  the total
event rate. (events/day)(dashed curve = no-oscillations)} 
\end{figure}

At the higher end of the mass range, as in fig.(4),
the difference in
seasonal modulation at different energies is noticeable. This is
true even when the integrated event rate variation does not differ
significantly from that expected for no oscillations.  This unique and
unambiguous signature of just-so oscillations means that measurements of 
the seasonal variation of the differential spectrum can be of great utility
when combined those of the overall event rate. Note that if the
Super-K energy resolution ($\sigma \approx 0.58 \sqrt{E}$ MeV)
could be sharpened, the above effect would be dramatically enhanced!     

Note (i.e. see \cite{GlashowKrauss}) 
that the seasonal variation in both the Cl and
Ga signals can be less than $10\%$ for just-so oscillations because of the
integration over energies inherent in 
chemical detectors (this effect is most striking for the Cl detector). 
While the Ga signal is expected to have a larger seasonal variation than Cl, it
may be small enough to have been missed.

%\vskip 3.8in
\begin{figure}[tbp]
\epsfxsize = \hsize \epsfbox{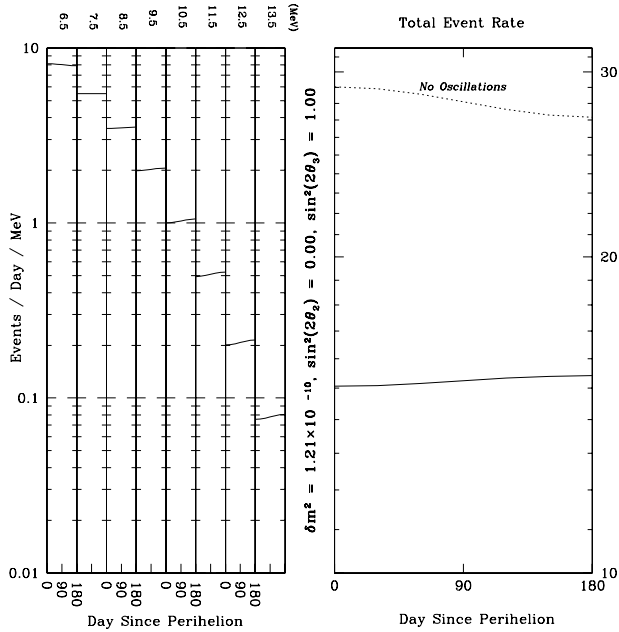}
\caption{ same as figure (2), for $\Delta m^2=1.2
\times10^{-10}$ eV$^2$, unfavored as described in the text}
\end{figure}

%\vskip 3.8in
\begin{figure}[tbp]
\epsfxsize = \hsize \epsfbox{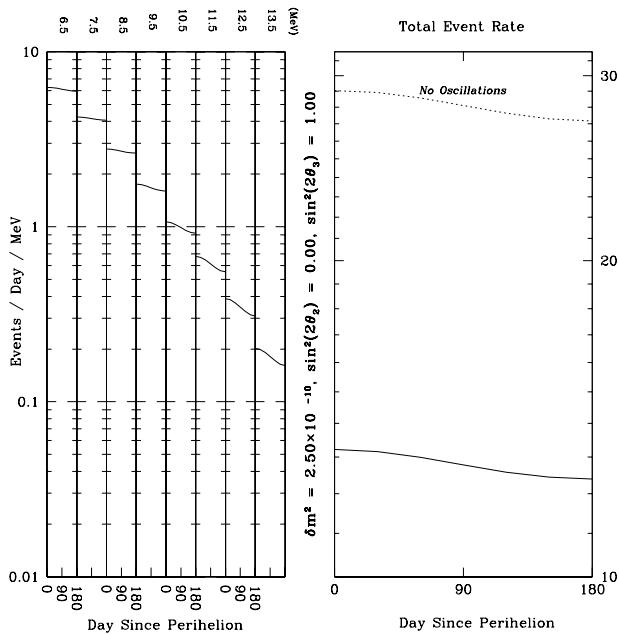}
\caption{ same as figure (2), for $\Delta m^2 =2.5\times10^{-10}$ eV$^2$.}
\end{figure}

Finally, 
a larger annual variation is expected in a detector such as
Borexino, which is strongly sensitive to the 0.86 MeV $^7$Be neutrino line.  
Seasonal variations as large as $40\%$ can be possible at this energy
\cite{GlashowKrauss,lisi}.  Perhaps even more exciting is the fortuitous
result
\cite{KraussWilczek} that the {\it phase} of time-dependent oscillations in the
Be neutrino signal, combined with the amplitude of the oscillations, can give a
direct measurement of the central solar core temperature.   While the
possibility of utilizing oscillations for this purpose exists only over a
limited range of masses and mixing angles in the just-so range, it would be
remarkable if nature allowed such a measurement.

The just-so solution of the solar
neutrino problem regained credibility as a result of recent
Super-K data. Furthermore, 
the technique that Super-K used to provide its
strongest evidence for  atmospheric neutrino oscillations---a time and
energy dependent modulation of the neutrino signal---may be exploited to
resolve one of the longest standing puzzles in particle astrophysics.

This work was supported in part by the DOE and the National Science
Foundation under grant number NSF-PHY/980-0729. LMK thanks the Aspen
Center for Physics, and A. Smirnov for useful comments.

\end{document}